\documentclass[fleqn,10pt]{wlscirep}
\usepackage{float}
\usepackage{multirow}
\usepackage[utf8]{inputenc}
\usepackage[T1]{fontenc}
\usepackage{booktabs}
\usepackage{xcolor}
\usepackage{amsmath}
\usepackage{graphicx}
\usepackage{subcaption}
\usepackage{siunitx}
\usepackage{physics}
\title{Improving Entanglement Resilience in Quantum Memories with Error-Detection-Based Distillation}

\author[1]{Huidan Zheng}
\author[1]{Gunsik Min}
\author[2]{Ilkwon Sohn}
\author[1,*]{Jun Heo}

\affil[1]{School of Electrical Engineering, Korea University, Seoul 02841, Republic of Korea}
\affil[2]{Quantum Network Research Center, Korea Institute of Science and Technology Information, Daejeon 34141, Republic of Korea}
\affil[*]{junheo@korea.ac.kr}

\begin{abstract}
The degradation of entanglement in quantum memories due to decoherence is a critical challenge for scalable quantum networks. We present an entanglement distillation protocol based on the $[[4,2,2]]$ quantum error-detecting code, deriving analytical expressions for its output fidelity and yield, and benchmarking it against the BBPSSW protocol. In addition to single-round performance, we further examine the iterative behavior of both protocols through multi-round fidelity and cumulative yield analysis. We then investigate a storage-stage recovery strategy in which the retained logical entangled state is subjected to repeated stabilizer-based syndrome checks without decoding and re-encoding, avoiding the need to regenerate and redistribute entanglement from scratch. Our analysis shows that this strategy can extend the usable storage lifetime beyond the BBPSSW baseline when the classical communication latency is sufficiently small. We derive latency thresholds and quantify the cumulative acceptance probabilities for different numbers of syndrome checks. 
A sensitivity analysis further shows that local gate and measurement errors reduce the fidelity advantage region and the admissible communication latency window, highlighting the importance of sufficiently accurate local operations. These results provide a quantitative framework for assessing the storage-stage benefit of logical state retention in the presence of finite communication latency and nonideal local operations.
\end{abstract}
\begin{document}

\flushbottom
\maketitle
%
%
\thispagestyle{empty}


\section*{Introduction}

Entanglement\cite{horodecki2009quantum} is a unique and essential resource in quantum information science. Its distinctive properties enable applications that are impossible with classical communication, such as quantum teleportation for transferring quantum states over arbitrary distances \cite{bennett1993teleporting} and quantum key distribution for establishing unconditionally secure communication channels \cite{bennett2014quantum}. In quantum network architectures, many quantum applications critically depend on entanglement, making entanglement-assisted networks a dominant paradigm in which entanglement itself is the central component\cite{li2023entanglement}.

However, the practical use of entanglement is fraught with challenges. Entangled states are inherently fragile and prone to degradation from various external influences, including channel noise during transmission and unavoidable interactions with the surrounding environment\cite{tanimura2014quantum}. These factors cause a gradual loss of entanglement, or decoherence\cite{zurek2003decoherence}, which reduces the quality of the entangled states and ultimately limits their usefulness for reliable quantum operations.

To counteract the degradation of entangled states under the constraint imposed by the no-cloning theorem, two primary strategies have been developed:  entanglement distillation protocols (EDPs)\cite{bennett1996purification} and quantum error correction (QEC)\cite{shor1995scheme, steane1996error}. EDPs typically consume multiple pairs of low-fidelity entangled states to probabilistically generate a smaller ensemble of entangled pairs with significantly higher fidelity, achieved through local operations and classical communication (LOCC). In contrast, QEC employs quantum error correction codes to encode quantum information across multiple physical qubits, enabling the detection and subsequent correction of errors. The relationship between QEC and EDP has been explored in previous theoretical studies\cite{bennett1996mixedstate, aschauer2005quantum}.

Although the main goal of entanglement distillation is to produce high-fidelity entangled pairs, a practical challenge arises when these distilled pairs are stored in quantum memories\cite{simon2010quantummemories} rather than used immediately. During storage, these states remain susceptible to decoherence, which can degrade their fidelity below the threshold required for reliable operations. In such cases, conventional EDPs\cite{bennett1996purification, deutsch1996quantum} may require discarding the degraded resources and restarting the entire process, an approach that is often inefficient\cite{bennett1996mixedstate}.

Previous studies have explored the use of quantum error-detecting or error-correcting codes in entanglement distillation, and a number of code-assisted schemes have been proposed to detect or correct errors in shared entangled pairs and thereby improve the quality of the distilled output. Beyond general code-based distillation approaches~\cite{aschauer2005quantum, wilde2010convolutional, dur2007entanglement, yu2025stabilizer, cheng2025adaptive, patil2024entanglement}, several studies have developed more specialized code-assisted frameworks for entangled-pair error detection, correction, and post-selection~\cite{Yumang2020QKD,Du2024advantage}. However, these works focus primarily on the initial distillation or error-filtering stage itself, rather than on the subsequent storage-stage behavior of the retained entangled resource under quantum memory decoherence. In this study, we instead focus on a storage-oriented scenario in which the post-selected output is retained as a logical entangled state in quantum memory and, when necessary, subjected to additional stabilizer-based syndrome checks without decoding and re-encoding.

\begin{figure}[ht]
\centering
\includegraphics[width=0.55\textwidth]{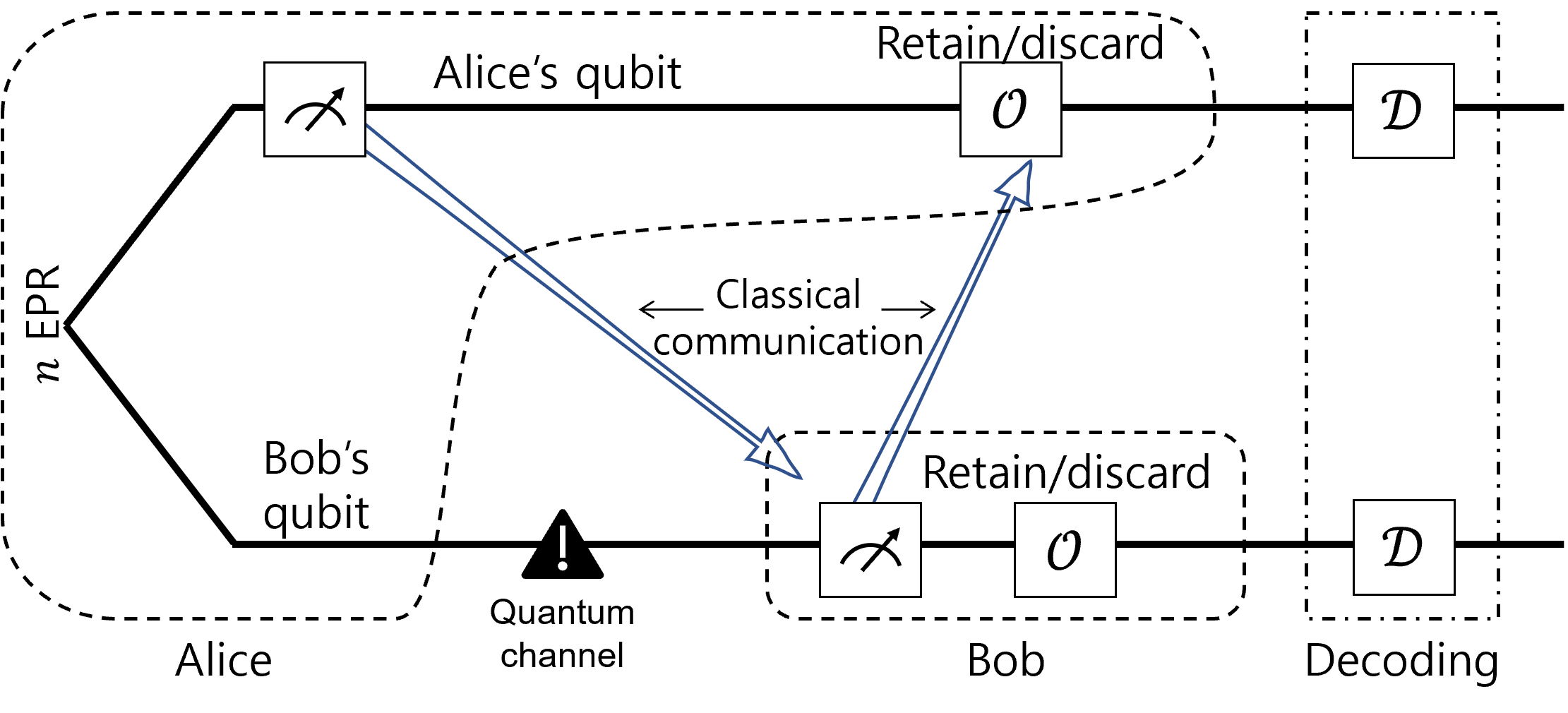}
\caption{Conceptual schematic of a probabilistic EDP.
An initial set of $n$ entangled pairs is distributed through a noisy quantum channel. 
Alice and Bob perform local operations and measurements, exchange outcomes via two-way classical communication,
and post-select by retaining or discarding pairs based on the results. Pairs that pass the parity check are retained and may subsequently be decoded for use. }
\label{fig:EDP using 2-way classical communication}
\end{figure}

The basic structure of this detection-based approach is illustrated in Fig.~\ref{fig:EDP using 2-way classical communication}. The protocol relies on local quantum operations, such as stabilizer measurements \cite{gottesman1997stabilizer}, to detect the presence of errors. The outcomes of these measurements determine whether the protocol passes the parity check and are communicated through two-way classical communication, which serves as a heralding mechanism that allows both parties to collaboratively decide whether to retain or discard the purified states.

However, a detailed analysis of how this EDC-based structure specifically benefits entangled states stored in quantum memory\cite{liu2023quantum}, along with the practical trade-offs involved, remains limited. 
To make this storage-stage behavior analytically explicit, we focus on the $[[4,2,2]]$ code  ~\cite{grassl1997codes} as a concrete case study, since it provides a simple nontrivial setting that captures the essential features of detection-based distillation, including post-selection, logical-state retention, and the trade-off between fidelity and yield. Within this framework, we derive analytic expressions for the output fidelity and yield, characterize the retained post-selected state as a logical entangled resource, and quantify the Bell-pair resource efficiency of reaching a target fidelity. As a benchmark, we compare the protocol with the BBPSSW scheme, which serves as a canonical recurrence-based baseline with analytically tractable performance metrics. We then examine the storage of the retained logical block under memory dephasing and repeated syndrome checking, explicitly accounting for post-selection probability and classical communication latency. We further assess the sensitivity of the resulting fidelity and latency advantages to local gate and measurement errors. This analysis characterizes the trade-offs among fidelity improvement, resource yield, storage lifetime, communication latency, and local operation quality.

For clarity, the baseline analyses below use the following assumptions: (i) ideal local quantum gates, (ii) ideal measurement outcomes, (iii) independent depolarizing noise on the transmitted qubits, (iv) independent pure dephasing of the physical qubits during quantum memory storage, and (v) Werner-state twirling between successive purification rounds in the iterative analysis. The finite syndrome-extraction time and classical communication latency are nevertheless included in the elapsed memory dephasing time. Operation-induced gate and measurement errors are treated separately in the supplementary sensitivity analysis.

\section*{Results}\label{sec:Results}


\subsection*{EDP using the $[[4,2,2]]$ quantum EDC}

Entanglement purification protocols improve the fidelity of entangled pairs by performing local operations on $n$ noisy pairs and using classical communication to decide, through post-selection, which pairs to retain. This process yields $m$ higher-fidelity pairs, with $n>m$. A key feature is the ability to discard or accept outcomes based on measurement results, allowing for repeated attempts at purification.

Unlike direct quantum state transmission, which risks total loss upon decoherence, these protocols help minimize overall fidelity loss. To further enhance this process, we incorporate a quantum EDC, which can identify outcomes for which no error is detected and inform retention decisions. Specifically, we consider the $[[4,2,2]]$-based protocol as a compact code-assisted distillation scheme and use BBPSSW as the benchmark protocol in the following performance comparison. The $[[4,2,2]]$-based EDP proceeds as follows:

\begin{figure}[ht]
\centering
\includegraphics[width=0.8\textwidth]{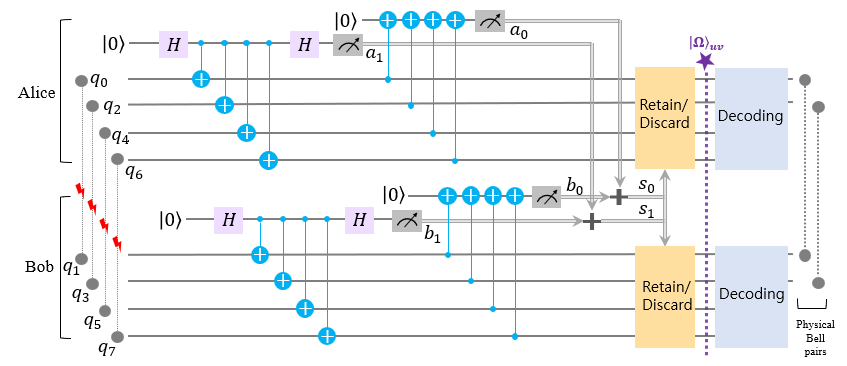}
\caption{Quantum circuit for the EDP using the $[[4,2,2]]$ code. Four entangled pairs are processed by Alice and Bob using local Hadamard gates, CNOT gates, and stabilizer measurements to prepare a syndrome-resolved logical entangled block. The measurement outcomes $a_0$ and $a_1$ denote Alice's two syndrome bits, while $b_0$ and $b_1$ denote Bob's corresponding two syndrome bits. From these, the relative parity check values $s_0=a_0\oplus b_0$ and $s_1=a_1\oplus b_1$ are obtained to determine whether the resulting block is retained or discarded. If retained, the corresponding syndrome-resolved logical entangled state is kept together with the common local syndrome, which is subsequently incorporated into the Pauli-frame description for storage or decoding.}
\label{fig:circuit1}
\end{figure}

\begin{enumerate}
    \item Alice prepares four entangled pairs, each initially in the Bell state $\lvert\Phi^+\rangle=\frac{1}{\sqrt{2}}(\lvert00\rangle+\lvert11\rangle)$.

    \item Alice performs the $[[4,2,2]]$ stabilizer measurements on her four qubits $(q_0,q_2,q_4,q_6)$ and records the two syndrome bits $a_0$ and $a_1$, as shown in Fig.~\ref{fig:circuit1}.

    \item Alice then sends the other halves of the four entangled pairs to Bob, with each transmitted qubit independently undergoing a depolarizing channel. For convenience in the subsequent error-pattern analysis, we use the Pauli-form parametrization
    \begin{equation}
    \mathcal{E}(\rho)
    =(1-p)\rho
    +\frac{p}{3}\left(X\rho X+Y\rho Y+Z\rho Z\right),
    \label{eq:depolarizing channel}
    \end{equation}
    where $p$ denotes the total probability of a nontrivial Pauli error and $\rho$ is the density matrix of the quantum state. This parametrization is equivalent to the more standard depolarizing form~\cite{Nielsen2010Quantum} $\mathcal{D}_q(\rho)=(1-q)\rho+q\frac{I}{2}$, with the parameter relation $q=4p/3$.

    \item Bob performs the same $[[4,2,2]]$ stabilizer measurements on his four qubits $(q_1,q_3,q_5,q_7)$ and records the corresponding syndrome bits $b_0$ and $b_1$.

    \item Alice and Bob compare the syndrome bits $(a_0,a_1)$ and $(b_0,b_1)$ through classical communication. They compute the relative parity-check values $s_0=a_0\oplus b_0$ and $s_1=a_1\oplus b_1$, which determine whether the state passes the protocol's post-selection criterion.

    \item If $(s_0,s_1)=(0,0)$, equivalently $a_0=b_0$ and $a_1=b_1$, Alice and Bob retain the corresponding syndrome-resolved logical entangled state. Otherwise, the state is discarded. The common local syndrome itself is not required to be $(0,0)$.

    \item For a retained state, Alice and Bob preserve the common local syndrome as classical side information. The corresponding syndrome sector is subsequently handled through the Pauli-frame convention defined below. When required, Alice and Bob decode their local $[[4,2,2]]$ blocks to obtain two physical Bell-pair outputs.
\end{enumerate}

Conditioned on a matched pair of local stabilizer outcomes, the retained state belongs to one of four syndrome-resolved logical entangled sectors. For the ideal accepted component, i.e., in the absence of an undetected channel error, let $u,v\in\{0,1\}$ denote the common local stabilizer-outcome bits associated with the $X^{\otimes4}$ and $Z^{\otimes4}$ measurements, respectively, where the corresponding eigenvalues are $(-1)^u$ and $(-1)^v$. The conditional state at the purple dashed line in Fig.~\ref{fig:circuit1} can then be written compactly as
\begin{equation}
\lvert\Omega_{uv}\rangle
=
\frac{1}{2}\Big[
\lvert\Phi^+\rangle^{\otimes4}
+(-1)^u\lvert\Psi^+\rangle^{\otimes4}
+(-1)^v\lvert\Phi^-\rangle^{\otimes4}
+(-1)^{u+v}\lvert\Psi^-\rangle^{\otimes4}
\Big],
\label{eq:logical_sector_422}
\end{equation}
where $\lvert\Phi^\pm\rangle=\frac{1}{\sqrt{2}}(\lvert00\rangle\pm\lvert11\rangle)$ and $\lvert\Psi^\pm\rangle=\frac{1}{\sqrt{2}}(\lvert01\rangle\pm\lvert10\rangle)$. The tensor products in Eq.~\eqref{eq:logical_sector_422} follow the four physical pairs $(q_0q_1)$, $(q_2q_3)$, $(q_4q_5)$, and $(q_6q_7)$.

Here, the $[[4,2,2]]$ structure is not used to encode an independently prepared logical input state. Instead, each local four-qubit block consists of one half of each of the four shared Bell pairs, and the local stabilizer measurements project the bipartite Bell-pair resource into syndrome-resolved sectors characterized by the $X^{\otimes 4}$ and $Z^{\otimes 4}$ stabilizers. We therefore refer to the retained state as a syndrome-resolved logical entangled block.

Acceptance requires only that Alice and Bob obtain the same local stabilizer outcomes, whereas the common value $(u,v)$ may be any of the four possible outcomes. The four accepted sectors are related by known bilateral Pauli transformations. We therefore retain $(u,v)$ as classical Pauli-frame information and represent each accepted block relative to the canonical sector $\lvert\Omega_{00}\rangle$, without applying additional physical correction gates. This canonical Pauli-frame convention is used throughout the subsequent storage analysis. After inverse $[[4,2,2]]$ decoding, the ideal canonical state yields two physical $\lvert\Phi^+\rangle$ Bell pairs. The explicit conditional states, Pauli-frame relations, and decoding procedure are derived in the Supplementary Information.

\begin{table}[H]
\centering
\caption{Outcome interpretation based on parity checks}
\label{tab:decision}
\begin{tabular}{ccc}
\toprule
$s_0=a_0\oplus b_0$ & $s_1=a_1\oplus b_1$ & Implication/Outcome \\
\midrule
0 & 0 & Pass (no detected error)\\
0 & 1 & Fail (possible $Z$ error)\\
1 & 0 & Fail (possible $X$ error)\\
1 & 1 & Fail (possible $Y$ error, or $X$ and $Z$ errors on different entanglements)\\
\bottomrule
\end{tabular}
\end{table}

Table~\ref{tab:decision}  summarizes the retain/discard decision based on the relative syndromes. Importantly, $(s_0,s_1)=(0,0)$ signifies agreement between Alice's and Bob's local syndromes; it does not require either local syndrome to be $(0,0)$. Here, ``no detected error'' means only that no error is indicated by the relative syndromes within the code's detection capability; undetectable error patterns may still pass the check and contribute to the residual output infidelity.

\subsection*{Performance comparison under output fidelity and yield between the
proposed EDC-based protocol and BBPSSW}

In this section, we compare the performance of the $[[4,2,2]]$ EDC-based EDP and the standard BBPSSW protocol, focusing on two key metrics: output fidelity and yield. In the protocol description above, the noisy transmission is parameterized by the depolarizing parameter \(p\). For the performance comparison in this section, however, it is more convenient to characterize the input state by the Bell-pair fidelity
\begin{equation}
F_{\mathrm{in}}=\langle \Phi^{+}|\rho_{\mathrm{in}}|\Phi^{+}\rangle,
\end{equation} where \(\rho_{\mathrm{in}}\) denotes the two-qubit state of a single raw Bell pair after one transmitted qubit has passed through the depolarizing channel. Here, $F_{\mathrm{in}}$ is introduced as a convenient input parameter for the analytical performance comparison and characterizes the quality of each raw Bell pair. It is defined at the level of an individual input pair and is not intended to describe the full multi-qubit state at a particular intermediate stage of the protocol. Under the channel model of Eq.~\eqref{eq:depolarizing channel}, this input fidelity is given by $F_{\mathrm{in}}=1-p$.

To compare the two protocols on a common input-output basis, we treat the EDC-based protocol as a \(4\to2\) scheme and consider two independent \(2\to1\) BBPSSW processes operating on four input pairs. The single-copy fidelity of each successful BBPSSW output pair is therefore the standard one-round BBPSSW fidelity. The explicit standard expression for \(F_{\mathrm{out}}^{\mathrm{BBPSSW}}\), rewritten in terms of \(F_{\mathrm{in}}\), is given later in the Methods section.

Since the $[[4,2,2]]$-based protocol produces two decoded physical Bell pairs upon acceptance, the output fidelity must also be specified consistently. In this work, we use the standard single-copy fidelity of the decoded output state,
\begin{equation}
F_{\mathrm{out}}^{\mathrm{EDC}}
=
F_{\mathrm{out}}^{(1)}
=
\langle \Phi^{+}|
\rho_{\mathrm{out}}^{(1)}
|\Phi^{+}\rangle .
\label{eq:Fout_definition}
\end{equation}
The two decoded output slots are symmetric under the present error model, so \(\rho_{\mathrm{out}}^{(1)}\) may be taken as the reduced state of either decoded output pair, equivalently their average reduced state, conditioned on acceptance. In the present error-pattern enumeration, this quantity can be evaluated as
\begin{equation}
F_{\mathrm{out}}^{(1)}
=
\frac{p_{\mathrm{corr}}^{\mathrm{avg}}}
{p_{\mathrm{pass}}},
\label{eq:Fout_eval}
\end{equation}
where \(p_{\mathrm{pass}}\) is the total probability of accepted outcomes, and \(p_{\mathrm{corr}}^{\mathrm{avg}}\) is the probability-weighted average number of decoded output pairs in the target Bell state \(|\Phi^{+}\rangle\), normalized per output pair. Equivalently, for each accepted error pattern, the contribution is assigned as \(1\), \(1/2\), or \(0\) depending on whether two, one, or none of the decoded output pairs are in \(|\Phi^{+}\rangle\), respectively. A detailed derivation and explicit examples are provided in the Supplementary Information.

For the $[[4,2,2]]$ EDC-based protocol, the pass probability and output fidelity take the explicit forms
\begin{equation}
p_{\text{pass}}^{\text{EDC}}
=
F_{\text{in}}^4
+
18\cdot\frac{F_{\text{in}}^2(1-F_{\text{in}})^2}{9}
+
24\cdot\frac{F_{\text{in}}(1-F_{\text{in}})^3}{27}
+
21\cdot\frac{(1-F_{\text{in}})^4}{81},
\label{ppass_EDC}
\end{equation}
\begin{equation}
F_{\text{out}}^{\text{EDC}}
=
\frac{
F_{\text{in}}^4
+
4\cdot\frac{F_{\text{in}}^2(1-F_{\text{in}})^2}{9}
+
4\cdot F_{\text{in}}(1-F_{\text{in}})^3/27
+
7\cdot(1-F_{\text{in}})^4/81
}
{p_{\text{pass}}^{\text{EDC}}}.
\label{fidelity_EDC}
\end{equation}
Each term in Eqs.~\eqref{ppass_EDC} and \eqref{fidelity_EDC} corresponds to a specific class of error patterns and their associated probabilities. The numerical coefficients reflect the number of error patterns in each class identified through the stabilizer-based acceptance condition. Undetectable patterns that pass the parity check are included in $p_{\mathrm{pass}}$; their contributions to the fidelity numerator depend on how many decoded output pairs remain in $|\Phi^+\rangle$.

The single-copy fidelity in Eq.~\eqref{eq:Fout_definition} does not by itself specify whether both decoded output pairs are simultaneously in the target Bell state. We therefore also characterize the accepted two-pair output by the joint fidelity
\[
F_{\mathrm{out}}^{(2)}
=
\langle\Phi^{+}\Phi^{+}|
\rho_{\mathrm{out}}^{(2)}
|\Phi^{+}\Phi^{+}\rangle ,
\]
where \(\rho_{\mathrm{out}}^{(2)}\) is the joint state of the two decoded output pairs conditioned on acceptance. For the EDC-based protocol, the two decoded outputs are generally correlated, so this quantity cannot in general be obtained by squaring the single-copy fidelity. From the accepted error-pattern distribution,
\[
F_{\mathrm{out,EDC}}^{(2)}
=
\frac{
F_{\mathrm{in}}^4+(1-F_{\mathrm{in}})^4/27
}
{p_{\mathrm{pass}}^{\mathrm{EDC}}}.
\]
Thus, \(F_{\mathrm{out}}^{(1)}\) characterizes the marginal quality of either decoded pair, whereas \(F_{\mathrm{out}}^{(2)}\) characterizes simultaneous two-pair correctness. The target-fidelity comparisons below use \(F_{\mathrm{out}}^{(1)}\) as the fidelity criterion, while \(F_{\mathrm{out}}^{(2)}\) is reported separately as a complementary two-output metric.

Yield quantifies the fraction of surviving output pairs relative to the input:
\begin{equation}
Y=\frac{m}{n}p_{\mathrm{pass}},
\end{equation}
where \(m/n\) is the output-to-input ratio and \(p_{\mathrm{pass}}\) denotes the probability that the corresponding post-selection criterion is satisfied. For the $[[4,2,2]]$-based EDP,
\[
Y^{\mathrm{EDC}}(F_{\mathrm{in}})
=
\frac{1}{2}p_{\mathrm{pass}}^{\mathrm{EDC}}(F_{\mathrm{in}}).
\]
For BBPSSW, each independent \(2\to1\) process produces one output pair with probability \(p_{\mathrm{pass}}^{\mathrm{BBPSSW}}\). Two such processes operating on four input pairs therefore produce an expected \(2p_{\mathrm{pass}}^{\mathrm{BBPSSW}}\) output pairs, giving the pooled single-round yield
\begin{equation}
Y^{\mathrm{BBPSSW}}(F_{\mathrm{in}})
=
\frac{1}{2}p_{\mathrm{pass}}^{\mathrm{BBPSSW}}(F_{\mathrm{in}}).
\label{eq:BBPSSW_pooled_yield}
\end{equation}
This accounting retains successful BBPSSW outputs independently and does not require two parallel branches to succeed in the same attempt. For notational simplicity, we omit the explicit dependence of \(F_{\mathrm{out}}\), \(p_{\mathrm{pass}}\), and \(Y\) on \(F_{\mathrm{in}}\) when no ambiguity arises.

\begin{figure}[h]
\centering
\includegraphics[width=0.8\textwidth]{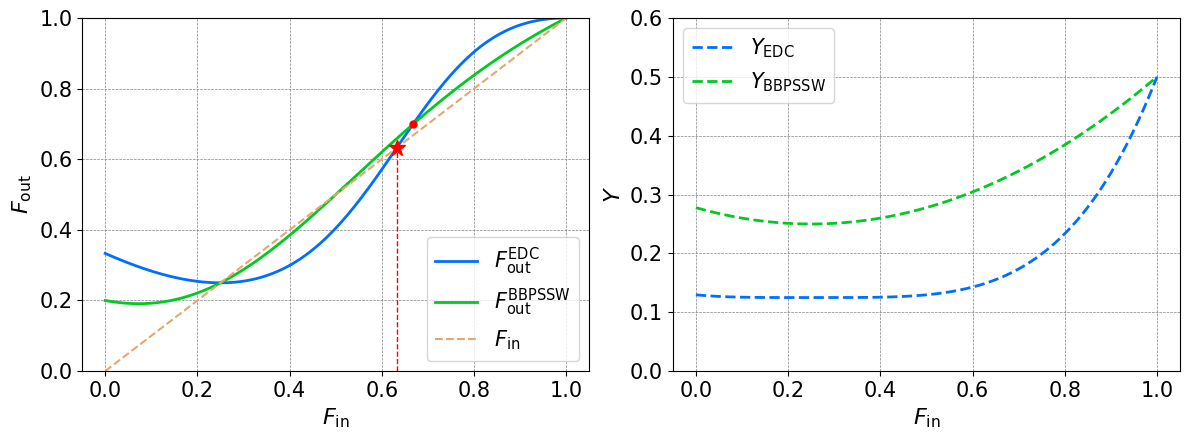}
\caption{Output fidelity $F_{\mathrm{out}}$ and yield $Y$ of the $[[4,2,2]]$ EDC-based EDP and BBPSSW protocol versus input fidelity $F_{\mathrm{in}}$. Solid curves show the single-copy output fidelities, and dashed lines indicate the corresponding single-round yields under pooled-output accounting. The red star marks the relevant purification threshold $F_{\mathrm{out}}^{\mathrm{EDC}}=F_{\mathrm{in}}$ in the entangled Werner-state regime at $F_{\mathrm{in}}\approx0.6323$, and the red dot marks the crossover at $F_{\mathrm{in}}\approx0.6675$, above which the EDC-based EDP has the higher single-copy output fidelity under the ideal local operation model.}
\label{fig:fidelity and yield}
\end{figure}

The comparative single-round results are shown in Fig.~\ref{fig:fidelity and yield}. Under the ideal local operation model used for this baseline comparison, the crossover at $F_{\mathrm{in}}\approx0.6675$ marks the point above which the $[[4,2,2]]$ EDC-based protocol provides a higher single-copy output fidelity than BBPSSW. The nontrivial solution of $F_{\mathrm{out}}^{\mathrm{EDC}}=F_{\mathrm{in}}$ for $F_{\mathrm{in}}>1/2$ occurs at $F_{\mathrm{in}}\approx0.6323$ and gives the relevant purification threshold in the entangled Werner-state regime. The single-round yield of the EDC-based protocol is lower than that of BBPSSW across the input-fidelity range under the pooled accounting used here. This reflects the stricter stabilizer-based post-selection of the EDC-based protocol, which rejects a broader set of error patterns while providing a larger fidelity gain above the crossover point.

We next examine how this fidelity--yield trade-off develops under repeated purification. Throughout the iterative analysis, a twirling operation is applied between successive distillation rounds, mapping the output state back to the Werner state family~\cite{Werner1989} before the subsequent round is initiated. This assumption ensures that the single-parameter fidelity-update function \(f(\cdot)\) remains well defined at each stage of the recursion: because \(F_{\mathrm{out}}^{\mathrm{EDC}}\) and \(F_{\mathrm{out}}^{\mathrm{BBPSSW}}\) are derived for Werner-state inputs, the iterative map \(F_{j+1}=f(F_j)\) is self-consistent when the inter-round output is restored to Werner form. The same twirling assumption is applied to both protocols.

Let \(F_0=F_{\mathrm{in}}\) and \(F_{j+1}=f(F_j)\), where \(f(\cdot)\) denotes the corresponding single-round fidelity-update function. The cumulative yield after \(r\) rounds is
\begin{equation}
Y_{\mathrm{cum}}^{(r)}
=
\prod_{j=0}^{r-1}Y(F_j),
\label{multiround yield}
\end{equation}
where \(Y(F_j)\) is the pooled single-round yield evaluated at the input fidelity of the \(j\)-th round. We use \(Y^{\mathrm{EDC}}(F_j)=\frac{1}{2}p_{\mathrm{pass}}^{\mathrm{EDC}}(F_j)\) and \(Y^{\mathrm{BBPSSW}}(F_j)=\frac{1}{2}p_{\mathrm{pass}}^{\mathrm{BBPSSW}}(F_j)\), consistent with the single-round comparison in Fig.~\ref{fig:fidelity and yield}. Here, \(Y_{\mathrm{cum}}^{(0)}=1\), and for \(r\geq1\), \(Y_{\mathrm{cum}}^{(r)}\) represents the expected surviving fraction of Bell-pair outputs relative to the initial input after \(r\) purification rounds.

\begin{figure}[H]
\centering
\includegraphics[width=0.85\textwidth]{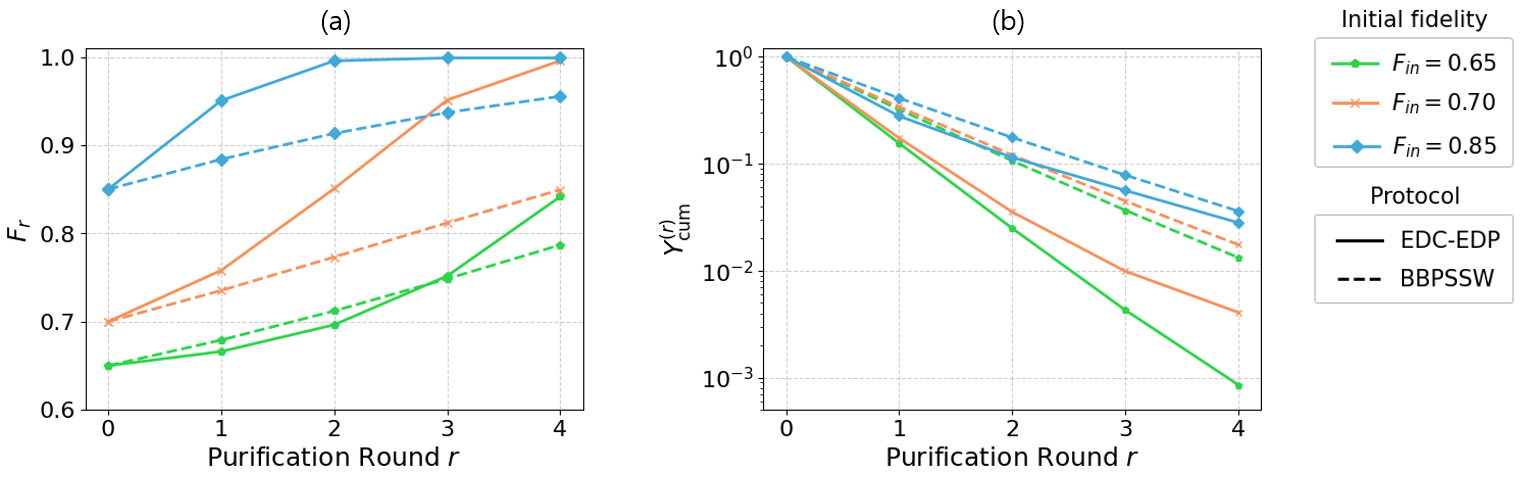}
\caption{Iterative performance comparison of the $[[4,2,2]]$-based EDP and the
BBPSSW protocol over multiple purification rounds. (a) Output fidelity $F_r$ as a
function of purification round $r$ for several initial input fidelities
$F_{\mathrm{in}}$. (b) Corresponding cumulative yield
$Y_{\mathrm{cum}}^{(r)}$, defined as the expected surviving fraction of Bell-pair
outputs relative to the initial input after $r$ recursive purification rounds.
Solid lines denote the $[[4,2,2]]$-based EDP, and dashed lines denote BBPSSW;
different colors correspond to different initial input fidelities. A twirling operation
is applied between successive rounds to restore the Werner-state form, ensuring
self-consistency of the iterative map $F_{j+1}=f(F_j)$ at each stage.}
\label{fig:output fidelity with 4 rounds}
\end{figure}

Fig.~\ref{fig:output fidelity with 4 rounds} shows the output fidelity $F_r$ and cumulative yield $Y_{\mathrm{cum}}^{(r)}$ for $r=0,1,2,3,$ and $4$. The same pooled single-round yield definition is used in both Figs.~\ref{fig:fidelity and yield} and \ref{fig:output fidelity with 4 rounds}, so the decrease in $Y_{\mathrm{cum}}^{(r)}$ directly reflects the successive resource reduction over multiple rounds.
 
 Under the same ideal local operation model, the single-round crossover at $F_{\mathrm{in}}\approx0.6675$ also shapes the iterative behavior. When the initial fidelity is below this point, as illustrated by \(F_{\mathrm{in}}=0.65\), the EDC-based EDP shows a weaker fidelity improvement than BBPSSW in the early rounds. Above the crossover, the EDC-based protocol provides a larger fidelity increase per round and can reach a given high-fidelity regime in fewer rounds. At the same round number, however, its cumulative yield remains lower because of its smaller single-round yield. The difference becomes less pronounced as the input states approach the target Bell state and the effect of the stricter EDC post-selection is reduced.

A target-oriented comparison must also account for the fact that the two protocols need not stop after the same number of rounds. For each representative \((F_{\mathrm{in}},F_{\mathrm{target}})\) setting, we define \(r\) as the minimum number of purification rounds required to satisfy the single-copy criterion
\[
F_{\mathrm{out}}^{(1)}\geq F_{\mathrm{target}}.
\]
The cumulative yield \(Y_{\mathrm{cum}}^{(r)}\) is then evaluated at this stopping round. The comparison assumes that two physical Bell-pair outputs are ultimately required; this specifies the output resource count, while \(F_{\mathrm{target}}\) remains a single-copy marginal-fidelity criterion.

For two independently generated BBPSSW outputs, the joint two-pair fidelity factorizes as
\[
F_{\mathrm{out,BBPSSW}}^{(2)}
=
\left(F_{\mathrm{out,BBPSSW}}^{(1)}\right)^2.
\]
For the EDC-based protocol, the two decoded outputs generally retain correlations from the accepted logical block, and their joint fidelity is evaluated directly from the accepted two-pair distribution as defined above.

\begin{table}[!htbp]
\centering
\caption{Target-oriented performance comparison under pooled resource accounting.
For each $(F_{\mathrm{in}},F_{\mathrm{target}})$ setting, $r$ is the minimum
number of purification rounds required to satisfy the single-copy criterion
$F_{\mathrm{out}}^{(1)}\geq F_{\mathrm{target}}$.
$F_{\mathrm{out}}^{(2)}$ denotes the joint two-pair fidelity of the resulting
outputs, and $Y_{\mathrm{cum}}^{(r)}$ is the cumulative yield evaluated at the
corresponding stopping round.}
\label{tab:table2}
\begin{tabular}{cccccc}
\toprule
$(F_{\mathrm{in}},F_{\mathrm{target}})$
& Protocol
& $r$
& $F_{\mathrm{out}}^{(1)}$
& $F_{\mathrm{out}}^{(2)}$
& $Y_{\mathrm{cum}}^{(r)}$ \\
\midrule
\multirow{2}{*}{$(0.80,0.90)$}
& BBPSSW       & 3 & 0.9045 & 0.8182 & 0.0656 \\
& EDC-based EDP & 1 & 0.9040 & 0.8774 & 0.2335 \\
\midrule
\multirow{2}{*}{$(0.85,0.90)$}
& BBPSSW       & 2 & 0.9134 & 0.8343 & 0.1758 \\
& EDC-based EDP & 1 & 0.9506 & 0.9369 & 0.2786 \\
\midrule
\multirow{2}{*}{$(0.90,0.95)$}
& BBPSSW       & 3 & 0.9629 & 0.9272 & 0.0925 \\
& EDC-based EDP & 1 & 0.9803 & 0.9747 & 0.3366 \\
\bottomrule
\end{tabular}
\end{table}

Table~\ref{tab:table2} shows that the EDC-based protocol reaches each of the three representative single-copy targets in one round, whereas BBPSSW requires two or three rounds. Although BBPSSW has the larger single-round yield in Fig.~\ref{fig:fidelity and yield}(b), the additional rounds required to reach the same target fidelity successively reduce its cumulative yield. The EDC-based protocol therefore has the larger \(Y_{\mathrm{cum}}^{(r)}\) at the respective stopping rounds for the three cases considered here.

The joint fidelities are lower than the corresponding single-copy fidelities because the joint metric requires both output pairs to be simultaneously in the target Bell state. For example, at $(F_{\mathrm{in}},F_{\mathrm{target}})=(0.80,0.90)$, the EDC-based protocol satisfies the adopted single-copy criterion with $F_{\mathrm{out}}^{(1)}=0.9040$, while its joint two-pair fidelity is $F_{\mathrm{out}}^{(2)}=0.8774$.

\subsection*{Fidelity decay and storage-stage recovery in quantum memory considering classical communication latency}

While the initial distillation process is essential for generating high-fidelity entangled resources, an important consideration arises when the retained states are stored in quantum memory prior to use. During storage, decoherence continuously degrades the entangled resource and may eventually reduce its fidelity below the level required by the target application. Classical communication latency also constrains storage-stage recovery in quantum repeater architectures ~\cite{munro2015inside}. We therefore examine whether the logical entangled state retained by the $[[4,2,2]]$-based protocol can maintain usable fidelity for a longer storage interval through repeated stabilizer-based post-selection.

We use the same three $(F_{\mathrm{in}},F_{\mathrm{target}})$ settings listed in Table~\ref{tab:table2} for the storage analysis. The stored logical block is ultimately decoded into two symmetric physical Bell-pair outputs. Consistent with the performance comparison above, $F_{\mathrm{target}}$ denotes the decoded single-copy marginal fidelity threshold for either output pair; it is not a joint two-pair fidelity threshold.
The storage noise is restricted to independent pure dephasing of the physical memory qubits. For a single qubit, the corresponding phase-flip probability after a storage interval $t$ is
\[
q(t)=\frac{1-e^{-\Gamma t}}{2},
\]
where $\Gamma$ is the single-qubit dephasing rate.

Unlike the initial distillation analysis, the accepted $[[4,2,2]]$ output entering memory is generally not described by a single fidelity parameter. For finite $F_{\rm in}$, undetected transmission errors produce a correlated logical Pauli-diagonal state. After incorporating the accepted common syndrome into the Pauli frame, we write the stored state as
\begin{equation}
\rho_L(0)
=
\sum_{\mathbf{a},\mathbf{b}}
\lambda_{\mathbf{a},\mathbf{b}}
|\Omega_{\mathbf{a},\mathbf{b}}\rangle
\langle\Omega_{\mathbf{a},\mathbf{b}}|,
\label{eq:logical_sector_state}
\end{equation}
where $\mathbf{a},\mathbf{b}\in\{0,1\}^2$ label the logical $X$- and $Z$-Pauli sectors, respectively, relative to the canonical state $|\Omega_{00}\rangle$. The coefficients $\lambda_{\mathbf{a},\mathbf{b}}$ are obtained from the accepted physical Pauli error patterns. The average decoded single-copy fidelity of this state is $F_0=F_{\rm out}^{\rm EDC}$, consistent with Eq.~(4), but the complete sector distribution is retained when evaluating the subsequent storage dynamics.

In the storage stage, recovery is therefore not modeled by reapplying the original $4\rightarrow2$ fidelity map to $F_0$. Instead, the same retained eight-qubit block remains in memory without decoding or re-encoding. After a finite storage interval, Alice and Bob repeat their local $X^{\otimes4}$ and $Z^{\otimes4}$ stabilizer measurements using ancillary qubits and retain the block only when the newly recorded local syndromes match. We refer to this operation as a \emph{syndrome check}.

Under the dephasing-only model, a syndrome check associated with an elapsed interval $\tau$ is accepted with probability
\begin{equation}
p_{\rm chk}(\tau)
=
\frac{1+e^{-8\Gamma\tau}}{2},
\qquad
\mu(\tau)
=
\operatorname{sech}(4\Gamma\tau),
\label{eq:check_map}
\end{equation}
where $\mu(\tau)$ characterizes the contraction of the conditional logical phase-sector distribution. Let $A$ denote the marginal probability that a given decoded logical pair has no logical-$X$ Pauli-error component; by symmetry, the value is the same for both pairs. This quantity is fixed by the initial accepted logical-sector distribution and remains invariant under pure dephasing. Conditioned on $k$ successful syndrome checks separated by elapsed intervals $\tau_1,\ldots,\tau_k$, the average decoded single-copy fidelity and the cumulative check-acceptance probability are
\begin{equation}
F_k
=
\frac{A}{2}
+
\left(
F_0-\frac{A}{2}
\right)
\prod_{j=1}^{k}\mu(\tau_j),
\qquad
P_k
=
\prod_{j=1}^{k}p_{\rm chk}(\tau_j).
\label{eq:repeated_checks}
\end{equation}
In particular, $p_{\rm chk}(0)=\mu(0)=1$, so an immediately repeated stabilizer measurement does not itself increase the fidelity. The benefit of repeated checking instead arises from post-selecting detectable errors accumulated during finite storage intervals. Accordingly, the syndrome check acts as conditional error filtering rather than active correction: it suppresses the fidelity loss relative to unconditioned storage but does not restore the retained state to its initial fidelity $F_0$.

\begin{figure}[t]
    \centering
    \includegraphics[width=0.8\textwidth]{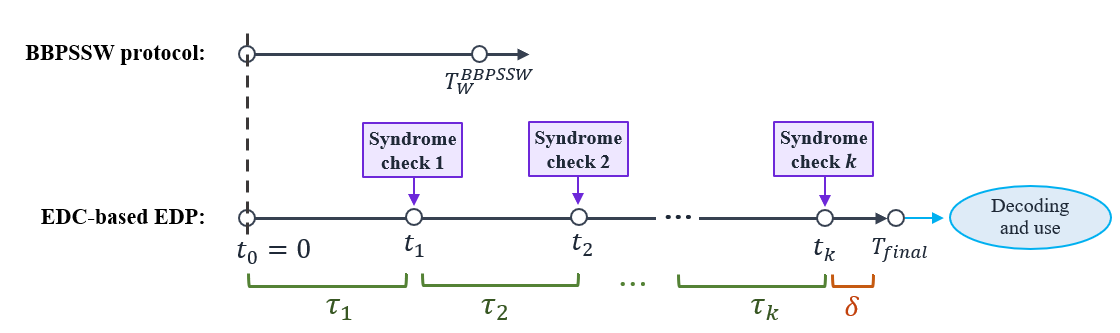}
    \caption{
    Measurement-centered timing diagram for storage and recovery.
    The upper line shows the storage timeline for a BBPSSW-prepared physical Bell pair, which remains directly usable up to the threshold time $T_W^{\rm BBPSSW}$.
    The lower line shows the corresponding timeline for the logical entangled state prepared by the $[[4,2,2]]$-based protocol.
    The recovery sequence is referenced to the completion times $t_j$ of successive local syndrome measurements, with $\tau_j=t_j-t_{j-1}$ denoting the elapsed dephasing interval between successive completed measurements.
    After the final syndrome measurement at $t_k$, the recorded outcomes are classically compared during the terminal interval $\delta=T_{\rm final}-t_k$.
    At $T_{\rm final}$, the final acceptance decision is available and an accepted logical block can proceed to decoding and use.
    The decoding operation itself is not included in the storage-and-recovery timing budget.
    }
    \label{fig:storage_timing}
\end{figure}

We describe the recovery sequence using the completion times of the local syndrome measurements as timing reference points. Let $t_0=0$ denote the time at which the initially accepted logical block enters memory, and let $t_j$ denote the completion time of the local syndrome-measurement step of the $j$th check. The corresponding elapsed dephasing interval is
\[
\tau_j=t_j-t_{j-1}.
\]
The first interval $\tau_1$ covers the memory exposure from $t_0$ to the completion of the first syndrome measurement, whereas $\tau_j$ for $j\ge2$ denotes the exposure between successive completed syndrome measurements. These intervals include the corresponding idle storage time and local syndrome-extraction duration; for $j\ge2$, they also include the classical communication delay following the preceding check.

After the $k$th local syndrome measurement is completed at $t_k$, its recorded syndrome information must still be classically compared. We define $T_{\rm final}$ as the time at which this final syndrome comparison is completed. At $T_{\rm final}$, the accept/reject decision and the corresponding Pauli frame are known, and an accepted logical block is ready for subsequent local decoding and use. The terminal interval
\[
\delta=T_{\rm final}-t_k,
\]
is not followed by another syndrome measurement. Consequently, dephasing accumulated during $\delta$ cannot be post-selected by the syndrome recorded at $t_k$ and therefore contributes as an unconditioned terminal dephasing interval. The resulting decoded average single-copy fidelity is
\begin{equation}
F_{\rm final}^{(k)}
=
\frac{A}{2}
+
e^{-4\Gamma\delta}
\left(
F_0-\frac{A}{2}
\right)
\prod_{j=1}^{k}\mu(\tau_j).
\label{eq:final_fidelity}
\end{equation}
The probability of obtaining this conditional output remains $P_k$ in Eq.~\eqref{eq:repeated_checks}, because the final classical comparison uses syndrome outcomes already recorded at $t_k$ and no additional syndrome measurement is performed during or after $\delta$. Equation~\eqref{eq:final_fidelity} uses the same decoded single-copy fidelity convention as Eq.~(4), allowing $F_{\rm final}^{(k)}$ to be compared directly with $F_{\rm target}$. Detailed logical-sector derivations of Eqs.~\eqref{eq:logical_sector_state}--\eqref{eq:final_fidelity} are provided in the Supplementary Information.

For the timing analysis, $T_{\rm cc}$ denotes the one-way classical communication latency, which is primarily determined by the separation and communication link between the remote nodes \cite{muralidharan2016optimal, wehner2018quantum}. We adopt a sequential round-trip classical-control convention in which the syndrome comparison and confirmation associated with a completed check are resolved before the next syndrome check is initiated. We model each such classical-control stage as $2T_{\rm cc}$. The local syndrome-check duration is represented by an effective processing time $T_{\mathrm{syn}}$, defined as
\begin{equation}
T_{\rm syn}
=
2T_s+4T_d+T_m,
\qquad
\delta=2T_{\rm cc},
\label{eq:timing_parameters}
\end{equation}
where $T_s$, $T_d$, and $T_m$ denote the durations of single-qubit gates, two-qubit gates, and measurements, respectively. 
$T_{\mathrm{syn}}$ is used as a coarse-grained effective local processing timescale rather than as a gate-by-gate circuit-depth count of the schematic implementation in Fig.~\ref{fig:circuit1}.
During this finite local processing time, the data qubits remain subject to the same memory dephasing process. Operation-induced two-qubit gate and measurement errors are not included in this baseline timing model; their effects on the fidelity crossover and latency thresholds are examined separately in the Supplementary Information using a simplified local noise model. The detailed interleaving of dephasing with individual gate layers is not modeled.

For the numerical storage analysis, we consider a uniform periodic-check schedule. For a fixed number of checks $k$, the interval from $t_0$ to the first completed syndrome measurement and the intervals between successive completed syndrome measurements are set equal:
\begin{equation}
\tau_1=\tau_2=\cdots=\tau_k\equiv\tau^{(k)}.
\end{equation}
Here, $\tau^{(k)}$ is determined separately for each value of $k$; thus, the intervals used for $k=1$, $2$, and $3$ are not required to be equal to one another. Physical timing feasibility requires
\[
\tau^{(1)}\ge T_{\rm syn},
\qquad
\tau^{(k)}\ge 2T_{\rm cc}+T_{\rm syn}
\quad (k\ge2).
\]
For $k\ge2$, although the first interval does not contain a preceding classical comparison delay, the common interval $\tau^{(k)}$ must be long enough to accommodate the mandatory $2T_{\rm cc}+T_{\rm syn}$ duration of each subsequent interval.

The final local decoding operation is not included in the storage-and-recovery timing budget. It is treated as an ideal local inverse-encoding operation performed after $T_{\rm final}$. This choice avoids introducing an implementation-specific decoding duration, since the circuit depth of the $[[4,2,2]]$ inverse encoder depends on the chosen gate decomposition and hardware architecture. 
Including a finite decoding time would introduce an additional unfiltered dephasing interval and would therefore shift the reported timing bounds downward without changing the structure of the present analysis. Syndrome-dependent Pauli corrections are likewise implemented through classical Pauli-frame tracking and require no additional physical correction-gate time.

Under the uniform-check schedule, the elapsed memory residency time from $t_0$ until the final syndrome-comparison result becomes available is
\begin{equation}
T_{\rm final}^{(k)}
=
k\tau^{(k)}+2T_{\rm cc}.
\label{eq:Tfinal_uniform}
\end{equation}
For a given $k$ and $T_{\rm cc}$, we vary the common interval $\tau^{(k)}$ and define the maximum usable memory residency time as
\begin{equation}
T_{\max}^{(k)}(T_{\rm cc})
=
\max_{\tau^{(k)}}
\left\{
k\tau^{(k)}+2T_{\rm cc}
\; \middle| \;
F_{\rm final}^{(k)}\ge F_{\rm target}
\ \text{and the timing constraints are satisfied}
\right\}.
\label{eq:Tmax}
\end{equation}
Thus, $T_{\rm final}^{(k)}$ denotes the elapsed time for a particular timing schedule, whereas $T_{\max}^{(k)}$ is the maximum value allowed by the target-fidelity and timing-feasibility conditions. Because $F_{\rm final}^{(k)}$ decreases monotonically with $\tau^{(k)}$ for fixed $k$ and $T_{\rm cc}$, the maximum is attained at $F_{\rm final}^{(k)}=F_{\rm target}$ whenever the corresponding timing-feasibility condition can be satisfied.

\begin{figure}[H]
\centering
\includegraphics[width=0.67\textwidth]{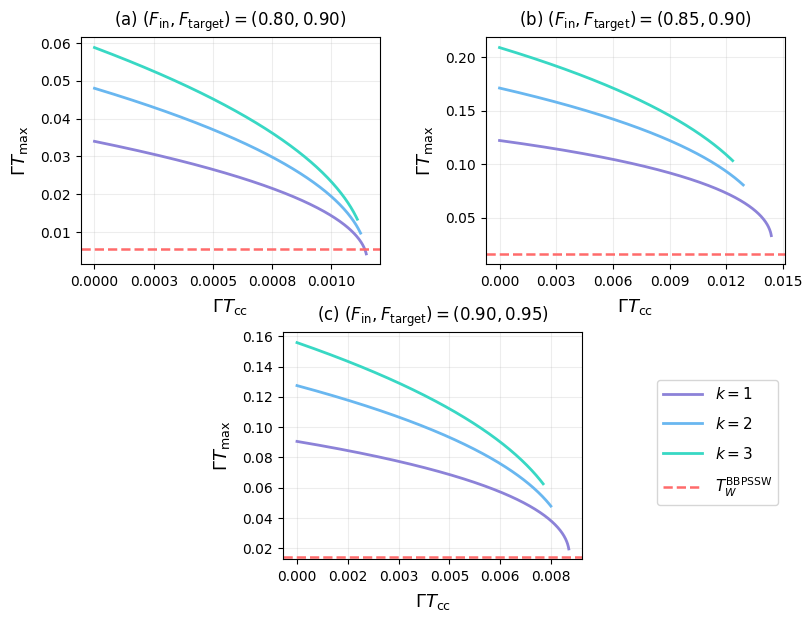}
\caption{
Maximum memory residency time of the stored entangled resource as a function of the normalized one-way classical communication latency for three scenarios:
    (a) $(F_{\rm in},F_{\rm target})=(0.8,0.9)$,
    (b) $(0.85,0.9)$, and
    (c) $(0.9,0.95)$.
    The curves labeled $k=1$, $2$, and $3$ show $\Gamma T_{\max}^{(k)}$ for one, two, and three uniformly spaced syndrome checks, respectively.
    For each $T_{\rm cc}$, $T_{\max}^{(k)}$ is the largest memory residency time for which the final conditional decoded fidelity remains at or above $F_{\rm target}$ while satisfying the timing-feasibility constraint.
    Each $k$-curve terminates when no timing-feasible uniform check interval can satisfy the target-fidelity condition.
    The dashed horizontal line denotes $\Gamma T^{\mathrm{BBPSSW}}_W$, the corresponding maximum storage time obtained from the Bell-diagonal output of the final BBPSSW round under the same pure-dephasing model, before its fidelity decreases to $F_{\mathrm{target}}$.
    }
\label{fig:repurification}
\end{figure}

All time quantities in the following numerical analysis are expressed in normalized units with respect to the memory coherence time,
\[
\Gamma t=\frac{t}{T_{\mathrm{coh}}},
\qquad
T_{\mathrm{coh}}=\frac{1}{\Gamma}.
\]
For the numerical examples, we adopt the normalized local time parameters
\[
T_s=2\times10^{-5},
\qquad
T_d=10^{-4},
\qquad
T_m=10^{-3},
\]
which are representative of the local gate and measurement time scales reported for contemporary quantum hardware~\cite{arute2019quantum, jurcevic2021demonstration, wright2019benchmarking}. This gives 
\[
T_{\rm syn}
=
2T_s+4T_d+T_m
=
1.44\times10^{-3}.
\]

For comparison with conventional physical-pair storage, we retain the full Bell-diagonal output distribution produced by the final BBPSSW round rather than replacing it by a Werner state before storage. 
If the final BBPSSW output has Bell-state populations 
$p_{\Phi^+}$, $p_{\Phi^-}$, $p_{\Psi^+}$, and $p_{\Psi^-}$, its fidelity with respect to $\ket{\Phi^+}$ under independent pure dephasing evolves as
\begin{equation}
F_{\mathrm{BBPSSW}}(t)
=
\frac{p_{\Phi^+}+p_{\Phi^-}}{2}
+
\frac{p_{\Phi^+}-p_{\Phi^-}}{2}
e^{-2\Gamma t}.
\label{eq:bbpssw-storage}
\end{equation}
We define $T_W^{\rm BBPSSW}$ as the time at which this fidelity decreases to $F_{\rm target}$. The Bell-state populations entering this expression and their dephasing evolution are derived explicitly in the Supplementary Information.

The EDC-based storage strategy retains a storage-lifetime advantage whenever
\[
T_{\max}^{(k)}(T_{\rm cc}) > T_W^{\rm BBPSSW}.
\]
The corresponding cumulative probability that the retained logical block passes all $k$ syndrome checks is given by $P_k$ in Eq.~\eqref{eq:repeated_checks}. This probability quantifies the probabilistic cost of repeated syndrome checking.

As shown in Fig.~\ref{fig:repurification}, increasing the number of syndrome checks extends the achievable memory residency time in the low-latency regime for all three scenarios, because additional checks provide further opportunities to filter detectable dephasing errors accumulated during storage. This benefit diminishes as $T_{\rm cc}$ increases, since the admissible check interval must simultaneously satisfy the target-fidelity and timing-feasibility conditions.

The termination of each curve marks the timing-feasibility limit. We define the $T_{\rm cc}$ threshold as the largest one-way classical communication latency for which a timing-feasible schedule still satisfies the storage-lifetime advantage criterion defined above. Table~\ref{tab:table3} summarizes these thresholds together with the cumulative syndrome-check acceptance probabilities $P_k$.
\begin{table}[h]
\centering
\caption{
Threshold comparison of classical communication latency for advantageous
storage-stage syndrome checking. The table lists the largest normalized
one-way classical communication latency for which each $k$-check strategy
remains timing-feasible and retains a storage-lifetime advantage over the
BBPSSW baseline, together with the corresponding cumulative syndrome-check
acceptance probability $P_k$.
}
\label{tab:table3}
\begin{tabular}{cccc}
\hline
$(F_{\rm in},  F_{\rm target})$& Strategy
& $\Gamma T_{\rm cc}$ threshold
& $P_k$ \\
\hline
$(0.80, 0.90)$& $k=1$ & 0.001140 & 0.9870 \\
& $k=2$ & 0.001124 & 0.9711 \\
& $k=3$ & 0.001111 & 0.9573 \\
\hline
$(0.85, 0.90)$& $k=1$ & 0.014407 & 0.9943 \\
& $k=2$ & 0.012923 & 0.8135 \\
& $k=3$ & 0.012360 & 0.7427 \\
\hline
$(0.90, 0.95)$& $k=1$ & 0.008035 & 0.9943 \\
& $k=2$ & 0.007497 & 0.8806 \\
& $k=3$ & 0.007271 & 0.8306 \\
\hline
\end{tabular}
\end{table}

The reported $P_k$ values refer only to the storage-stage syndrome checks in Eq.~\eqref{eq:repeated_checks} and are evaluated at the corresponding $T_{\rm cc}$ thresholds. Table~\ref{tab:table3} shows that the $T_{\rm cc}$ threshold decreases with increasing $k$ in all three scenarios, while the cumulative acceptance probability $P_k$ also decreases. Thus, the longer conditional memory residency time enabled by additional syndrome checks is accompanied by tighter classical communication requirements and a greater post-selection cost.

A supplementary sensitivity analysis including two-qubit gate and measurement errors shows that local operational noise further reduces the advantageous parameter region. For equal gate and measurement error rates of $10^{-3}$, the lower fidelity crossover shifts from approximately $0.6675$ to $0.6736$, while the admissible classical communication-latency windows are reduced, particularly when the initial EDC output lies only slightly above $F_{\mathrm{target}}$. The noise model and detailed crossover and latency-threshold results are provided in the Supplementary Information.

\section*{Discussion}


In this study, we analyzed an entanglement distillation protocol based on the $[[4,2,2]]$ error-detecting code and compared it with BBPSSW. Under the ideal local operation baseline, the $[[4,2,2]]$-based protocol provides a higher single-copy output fidelity above the crossover point, although its stricter post-selection results in a lower single-round yield. Under the target-oriented comparison with pooled BBPSSW output accounting, the EDC-based protocol reaches the three representative target fidelities considered here in fewer rounds and retains a larger cumulative yield at the corresponding stopping rounds. Because the two decoded EDC outputs are generally correlated, we also distinguish the single-copy fidelity from the joint two-pair fidelity. The retained post-selected state is further characterized explicitly as a syndrome-resolved logical entangled state, with the common local syndrome incorporated into a classical Pauli frame.

We then considered the storage of this retained logical state in quantum memory. Rather than reapplying the original $4\rightarrow2$ distillation map, the same retained logical block remains in memory and is periodically subjected to repeated local stabilizer measurements. The storage dynamics are evaluated from the full logical Pauli-sector distribution under physical-qubit dephasing. This treatment also accounts for the finite syndrome-extraction time, classical communication latency, the unfiltered dephasing accumulated after the final syndrome measurement, and the cumulative probability of successful syndrome checks. The results show that repeated syndrome checking can extend the usable storage time beyond the BBPSSW-prepared physical pair baseline when the communication latency is sufficiently small, at the cost of a reduced cumulative acceptance probability.

Our analysis is limited to a probabilistic error-detection-based protocol using two-way classical communication. An alternative approach would be to employ a full QEC code capable of active error correction, which may allow deterministic operation and reduce the classical communication requirement to one-way communication~\cite{munro2015inside}. Such an approach may be advantageous when communication latency is large relative to the memory coherence time, although the trade-off between communication requirements, fidelity improvement, and physical resources differs from that of probabilistic distillation protocols~\cite{aschauer2005quantum}. A systematic comparison of these two approaches is beyond the scope of the present work.

The main analytical baseline assumes ideal local gates and measurements. This assumption is particularly relevant to the EDC-based protocol because its stabilizer-extraction circuitry requires more local entangling operations than BBPSSW and is therefore more sensitive to operational errors. The supplementary sensitivity analysis confirms that imperfect local operations narrow both the fidelity-advantage region and the admissible classical communication latency window, with a stronger impact when the initial EDC output lies close to the target fidelity. These results are consistent with the known sensitivity of entanglement-purification and repeater protocols to imperfect local operations\cite{muralidharan2016optimal,dur1999quantum}. The ideal operation results should therefore be interpreted as applying to a regime of sufficiently accurate local control. A hardware-specific circuit-level noise treatment remains beyond the scope of this work. 

A further limitation concerns the input-fidelity regime below the ideal operation crossover at $F_{\mathrm{in}}\approx0.6675$, where the $[[4,2,2]]$-based protocol does not provide a single-round fidelity advantage over BBPSSW. The advantageous interval also changes when local operational errors are included. For lower-fidelity Werner-state inputs that remain within the BBPSSW purification regime, a hybrid strategy could first use BBPSSW to increase the fidelity before applying the EDC-based protocol for further purification and logical-state storage. Evaluating such a cascaded strategy would require accounting for its additional yield loss, local operations, and communication overhead.

Future work may extend the present analysis to other error-detecting or error-correcting codes and to hardware-specific implementations of the syndrome-extraction circuit. Experimental studies would also be useful for evaluating repeated syndrome checking under realistic memory and control errors. Overall, the results show that retaining the post-selected logical state can provide a storage-stage advantage without regenerating and redistributing entanglement from scratch, provided that classical communication is sufficiently fast and local operations are sufficiently accurate. This advantage is accompanied by lower single-round yield and by the probabilistic cost of repeated syndrome checking.

\section*{Methods}

\subsection*{Entanglement distillation protocols}

Entanglement distillation is a crucial technique for enhancing the fidelity of entangled pairs that have been degraded by noise, typically during distribution over imperfect quantum channels. One of the earliest and most widely referenced entanglement distillation protocols is the BBPSSW scheme, which has served as a foundational benchmark for subsequent developments. It is widely recognized for its conceptual simplicity and effectiveness, and its standard formulation assumes that the input states are Werner states, mixed states characterized by a single fidelity parameter with respect to the target Bell state. These features collectively motivate our selection of BBPSSW as the conventional benchmark for performance comparison. The detailed procedure of the BBPSSW scheme (two-copy purification with local CNOTs and measurement-based post-selection) is described in its original literature~\cite{bennett1996purification}.

Since its proposal, a variety of follow-up studies have explored related entanglement purification strategies and refinements under different state assumptions and operational settings ~\cite{deutsch1996quantum, Dehaene2004local, Hsieh2003A, Zhou2020Purification, maneva2000improved, dur2003entanglement}. In parallel, other classes of entanglement distillation protocols, such as hashing and breeding schemes~\cite{bennett1996mixedstate}, have been proposed to asymptotically extract high-fidelity entanglement from large ensembles of noisy states. However, these approaches generally require a substantial and uncertain number of low-fidelity entangled pairs, and the necessary input size $n$ is often difficult to estimate or optimize\cite{aschauer2005quantum}. Moreover, their decoding procedures can be computationally intensive and challenging to implement on current quantum hardware. These limitations make such protocols less suitable for resource-constrained scenarios and motivate the consideration of fixed-size, code-assisted protocols such as the one explored in this study.

In the BBPSSW protocol, pairs of entangled qubits undergo local CNOT operations followed by measurements and classical communication for post-selection, a sequence broadly similar to the operational structure of our EDC-based approach. The resulting distilled pair achieves an output fidelity
\begin{equation}
F_{\text{out}}^{\text{BBPSSW}} 
= \frac{F_{\text{in}}^2 + \frac{1}{9}(1-F_{\text{in}})^2}{p_{\text{pass}}^{\text{BBPSSW}}},
\end{equation}
where the pass probability of a single round is given by
\begin{equation}
p_{\text{pass}}^{\text{BBPSSW}}
= F_{\text{in}}^2 + \frac{2}{3}F_{\text{in}}(1-F_{\text{in}}) + \frac{5}{9}(1-F_{\text{in}})^2.
\end{equation}
These expressions serve as the foundation for the performance comparisons presented in the Results section.

\subsection*{Unitary properties of Bell states}

Understanding the unitary properties of Bell states is essential for explaining the feasibility and design rationale of our distillation protocol. Bell states exhibit a special linear algebraic property that plays a central role in the design of entanglement distillation and stabilizer-based protocols. When two parties, Alice and Bob, share a Bell state $|\Phi^+\rangle_{AB}=\frac{1}{\sqrt2}(|00\rangle_{AB}+|11\rangle_{AB})$, any linear operator applied to one qubit can be interpreted, up to a transpose, as acting on the other qubit\cite{ aschauer2005quantum}.

Formally, for any linear operator $U$ acting on Alice's side (subscripts $A$ and $B$ denote Alice’s and Bob’s qubits, respectively), the following identity holds\cite{bennett1996mixedstate, dur2007entanglement}:
\begin{equation}
(U_A\otimes I_B)|\Phi^+\rangle_{AB}=(I_A\otimes U_B^T)|\Phi^+\rangle_{AB}.
\end{equation}

This identity implies that applying a transformation to one half of a Bell pair induces a corresponding transformation on the other half, which makes Bell states particularly useful for distributed quantum operations such as quantum teleportation and parity checks in entanglement distillation.

This property generalizes to the case where multiple Bell pairs are used to construct a joint entangled state across $2n$ qubits. When local stabilizer projectors (e.g., parity checks) are applied on Alice's qubits, the matrix identity ensures that their action is mirrored onto Bob's side, effectively allowing global syndrome extraction through only local measurements. In the context of our protocol, where logical Bell states are constructed using multiple $|\Phi^+\rangle$ states, this duality enables us to implement multi-qubit stabilizer measurements like $X^{\otimes 4}$ and $Z^{\otimes 4}$ by performing local parity checks and sharing the outcomes over classical channels.

This matrix identity also provides an intuitive justification for why logical entangled states built from Bell pairs can be filtered using stabilizer-based measurements: the projection onto a stabilizer eigenspace on one side naturally induces the corresponding projection onto the stabilizer eigenspace on the other, preserving the desired symmetry structure of the logical state.

\section*{Funding}
H.Z. discloses support for the research of this work from the Institute of Information \& Communications Technology Planning \& Evaluation (IITP) grant funded by the Korean government (MSIT) [grant number 2022-0-00463]. 
I.S. discloses support for the research of this work from the Korea Institute of Science and Technology Information (KISTI) [grant number (KISTI)K26L1M3C5-01]. 
G.M. and J.H. disclose support for the research of this work from the IITP (Institute of Information \& Communications Technology Planning \& Evaluation)-ITRC (Information Technology Research Center) grant funded by the Korean government (Ministry of Science and ICT) [grant number IITP-2026-RS-2021-II211810].

\section*{Data availability}
The datasets used and/or analysed during the current study are available from the corresponding author on reasonable request.

\bibliography{references}

\section*{Acknowledgements}

This work was supported by the IITP (Institute of Information \& Communications Technology Planning \& Evaluation)-ITRC(Information Technology Research Center) grant funded by the Korea government(Ministry of Science and ICT)(IITP-2026-RS-2021-II211810).
 This work was supported by Institute of Information \& communications Technology Planning \& Evaluation (IITP) grant funded by the Korea government (MSIT) (No. 2022-0-00463, Development of a quantum repeater in optical fiber networks for quantum internet).
 This research was supported by Korea Institute of Science and Technology Information (KISTI). (No. (KISTI)K26L1M3C5-01).

\section*{Author contributions statement}


Huidan Zheng and Jun Heo developed the theoretical idea. Huidan Zheng performed the main analysis and should be regarded as the main author. Gunsik Min and Ilkwon Sohn provided helpful discussions and constructive feedback on the research and manuscript. All authors contributed to the discussion and interpretation of the results.

\section*{Additional information}



\textbf{Supplementary information} accompanies this paper.  \\
\textbf{Competing interests} The authors declare no competing interests.

\end{document}